\documentclass[twocolumn,showpacs,aps,prl,superscriptaddress]{revtex4}

\newcommand{\BaBarYear}       {08}
\newcommand{\BaBarNumber}     {022}
\newcommand{\SLACPubNumber} {13322}
\newcommand{\BaBarType}      {PUB}  

\long\def\inst#1{\par\nobreak\kern 4pt\nobreak
    {\it #1}\par\vskip 10pt plus 3pt minus 3pt}

\usepackage[dvips]{graphicx}
\usepackage{epsf}  
\usepackage{epsfig}  
\usepackage{rotating}
\usepackage{color}
\usepackage{colordvi}

\newcommand{\psfile}[3][]{ 
  \begin{center}
    \setlength{\epsfxsize}{#3\linewidth}\leavevmode
    \def\noOpt{}\def\testit{#1}\ifx\testit\noOpt%
      \epsfbox{#2}%
    \else%
      \epsfbox[#1]{#2}%
    \fi
  \end{center}
}

\RequirePackage{xspace}

\usepackage{relsize}
\def\babar{\mbox{\slshape B\kern-0.1em{\smaller A}\kern-0.1em
    B\kern-0.1em{\smaller A\kern-0.2em R}}}

\def\epem       {\ensuremath{e^+e^-}\xspace}

\def\mumu       {\ensuremath{\mu^+\mu^-}\xspace}

\def\ellell     {\ensuremath{\ell^+ \ell^-}\xspace}

\def\piz   {\ensuremath{\pi^0}\xspace}

\def\pip   {\ensuremath{\pi^+}\xspace}
\def\pim   {\ensuremath{\pi^-}\xspace}

\def\Kbar  {\kern 0.2em\overline{\kern -0.2em K}{}\xspace}

\def\Kz    {\ensuremath{K^0}\xspace}
\def\Kzb   {\ensuremath{\Kbar^0}\xspace}
\def\KzKzb {\ensuremath{\Kz \kern -0.16em \Kzb}\xspace}
\def\Kp    {\ensuremath{K^+}\xspace}
\def\Km    {\ensuremath{K^-}\xspace}

\def\KpKm  {\ensuremath{\Kp \kern -0.16em \Km}\xspace}
\def\KS    {\ensuremath{K^0_{\scriptscriptstyle S}}\xspace}

\def\Kstar   {\ensuremath{K^*}\xspace}

\def\Kstarp  {\ensuremath{K^{*+}}\xspace}

\def\Dbar    {\kern 0.2em\overline{\kern -0.2em D}{}\xspace}

\def\Dz      {\ensuremath{D^0}\xspace}
\def\Dzb     {\ensuremath{\Dbar^0}\xspace}
\def\DzDzb   {\ensuremath{\Dz {\kern -0.16em \Dzb}}\xspace}
\def\Dp      {\ensuremath{D^+}\xspace}
\def\Dm      {\ensuremath{D^-}\xspace}

\def\DpDm    {\ensuremath{\Dp {\kern -0.16em \Dm}}\xspace}

\def\B       {\ensuremath{B}\xspace}
\def\Bbar    {\kern 0.18em\overline{\kern -0.18em B}{}\xspace}

\def\BB      {\ensuremath{B\Bbar}\xspace}
\def\Bz      {\ensuremath{B^0}\xspace}
\def\Bzb     {\ensuremath{\Bbar^0}\xspace}
\def\BzBzb   {\ensuremath{\Bz {\kern -0.16em \Bzb}}\xspace}
\def\Bu      {\ensuremath{B^+}\xspace}
\def\Bub     {\ensuremath{B^-}\xspace}
\def\Bp      {\ensuremath{\Bu}\xspace}

\def\BpBm    {\ensuremath{\Bu {\kern -0.16em \Bub}}\xspace}

\def\BorBbar    {\kern 0.18em\optbar{\kern -0.18em B}{}\xspace}
\def\DorDbar    {\kern 0.18em\optbar{\kern -0.18em D}{}\xspace}
\def\KorKbar    {\kern 0.18em\optbar{\kern -0.18em K}{}\xspace}

\def\jpsi     {\ensuremath{{J\mskip -3mu/\mskip -2mu\psi\mskip 2mu}}\xspace}
\def\psitwos  {\ensuremath{\psi{(2S)}}\xspace}

\mathchardef\Upsilon="7107
\def\Y#1S{\ensuremath{\Upsilon{(#1S)}}\xspace}

\def\FourS {\Y4S}

\mathchardef\Deltares="7101
\mathchardef\Xi="7104
\mathchardef\Lambda="7103
\mathchardef\Sigma="7106
\mathchardef\Omega="710A

\def\Deltabar{\kern 0.25em\overline{\kern -0.25em \Deltares}{}\xspace}
\def\Lbar{\kern 0.2em\overline{\kern -0.2em\Lambda\kern 0.05em}\kern-0.05em{}\xspace}
\def\Sigbar{\kern 0.2em\overline{\kern -0.2em \Sigma}{}\xspace}
\def\Xibar{\kern 0.2em\overline{\kern -0.2em \Xi}{}\xspace}
\def\Obar{\kern 0.2em\overline{\kern -0.2em \Omega}{}\xspace}
\def\Nbar{\kern 0.2em\overline{\kern -0.2em N}{}\xspace}
\def\Xb{\kern 0.2em\overline{\kern -0.2em X}{}\xspace}

\def\mes        {\mbox{$m_{\rm ES}$}\xspace}

\def\DeltaE     {\mbox{$\Delta E$}\xspace}

\newcommand{\tev}{\ensuremath{\mathrm{\,Te\kern -0.1em V}}\xspace}
\newcommand{\gev}{\ensuremath{\mathrm{\,Ge\kern -0.1em V}}\xspace}
\newcommand{\mev}{\ensuremath{\mathrm{\,Me\kern -0.1em V}}\xspace}
\newcommand{\kev}{\ensuremath{\mathrm{\,ke\kern -0.1em V}}\xspace}
\newcommand{\ev}{\ensuremath{\mathrm{\,e\kern -0.1em V}}\xspace}
\newcommand{\gevc}{\ensuremath{{\mathrm{\,Ge\kern -0.1em V\!/}c}}\xspace}
\newcommand{\mevc}{\ensuremath{{\mathrm{\,Me\kern -0.1em V\!/}c}}\xspace}
\newcommand{\gevcc}{\ensuremath{{\mathrm{\,Ge\kern -0.1em V\!/}c^2}}\xspace}
\newcommand{\mevcc}{\ensuremath{{\mathrm{\,Me\kern -0.1em V\!/}c^2}}\xspace}

\def\mus  {\ensuremath{\rm \,\mus}\xspace}

\def\mus        {\ensuremath{\,\mu{\rm s}}\xspace}    

\def\to                 {\ensuremath{\rightarrow}\xspace}

\def\pep2{PEP-II}

\def\gsim{{~\raise.15em\hbox{$>$}\kern-.85em
          \lower.35em\hbox{$\sim$}~}\xspace}
\def\lsim{{~\raise.15em\hbox{$<$}\kern-.85em
          \lower.35em\hbox{$\sim$}~}\xspace}

\def\CP                {\ensuremath{C\!P}\xspace}

\xspace

\newcommand{\epjBase}        {Eur.\ Phys.\ Jour.\xspace}

\newcommand{\jprBase}        {Phys.\ Rev.\xspace}
\newcommand{\jplBase}        {Phys.\ Lett.\xspace}
\newcommand{\nimBaseA}       {Nucl.\ Instrum.\ Methods Phys.\ Res., Sect.\ A\xspace}

\newcommand{\nimBaseC}       {Nucl.\ Instrum.\ Methods Phys.\ Res., Sect.\ C\xspace}

\newcommand{\npBase}         {Nucl.\ Phys.\xspace}
\newcommand{\zpBase}         {Z.\ Phys.\xspace}

\newcommand{\epjc}      [1]  {\epjBase\ C~{\bf #1}}

\newcommand{\mpl}       [1]  {{Mod.\ Phys.\ Lett.\ {\bf #1}}}

\newcommand{\nim}       [1]  {\nimBaseC~{\bf #1}}

\newcommand{\nima}      [1]  {\nimBaseA~{\bf #1}}

\newcommand{\npb}       [1]  {\npBase\ B~{\bf #1}}

\newcommand{\npbps}     [1]  {{Nucl.\ Phys.\ B~Proc.\ Suppl.\ {\bf #1}}}

\newcommand{\plb}       [1]  {\jplBase\ B~{\bf #1}}

\newcommand{\pr}        [1]  {\jprBase\ {\bf #1}}

\newcommand{\progtp}    [1]  {{Prog.\ Theor.\ Phys.\ {\bf #1}}}

\newcommand{\zpc}       [1]  {\zpBase\ C~{\bf #1}}

\def\jetset74   {\mbox{\tt Jetset \hspace{-0.5em}7.\hspace{-0.2em}4}\xspace}

\newcommand{\gevcccc}{\ensuremath{{\mathrm{\,Ge\kern -0.1em V^2\!/}c^4}}\xspace}

\def\K {\ensuremath{K}\xspace}

\def\Kmaybestar {\ensuremath{K^{(*)}\xspace}}

\def\kll {\B\to\Kmaybestar\ellell\xspace}
\def\kllshort {(K,K^{*})\ellell\xspace}
\def\kmaybeee {\B\to\Kmaybestar\epem\xspace}

\def\kmaybemm {\B\to\Kmaybestar\mumu\xspace}

\def\modekavgll {\ensuremath{B\to K\ellell}\xspace}

\def\modekstee {\ensuremath{B\rightarrow K^{*}\epem}\xspace}

\def\modekstll {\ensuremath{B\rightarrow K^{*}\ellell}\xspace}

\def\modekllshort {\ensuremath{K^+\ellell}\xspace}
\def\modekavgmmshort {\ensuremath{K\mumu}\xspace}
\def\modekavgeeshort {\ensuremath{K\epem}\xspace}
\def\modekavgllshort {\ensuremath{K\ellell}\xspace}
\def\modekstkllshort {\ensuremath{K^{*0}\ellell}\xspace}
\def\modekstksllshort {\ensuremath{K^{*+}\ellell}\xspace}
\def\modekstmmshort {\ensuremath{K^{*}\mumu}\xspace}
\def\modeksteeshort {\ensuremath{K^{*}\epem}\xspace}
\def\modekstllshort {\ensuremath{K^{*}\ellell}\xspace}

\usepackage{relsize}
\usepackage{xspace}

\begin{document}


\noindent \babar-\BaBarType-\BaBarYear/\BaBarNumber \\
SLAC-PUB-\SLACPubNumber \\

\begin{flushright}
arXiv:0807.4119 \\
\end{flushright}

\title{
        {\mathversion{bold}
         Direct CP, Lepton Flavor and Isospin Asymmetries in the Decays $\kll$}
}

%
\author{B.~Aubert}
\author{M.~Bona}
\author{Y.~Karyotakis}
\author{J.~P.~Lees}
\author{V.~Poireau}
\author{E.~Prencipe}
\author{X.~Prudent}
\author{V.~Tisserand}
\affiliation{Laboratoire de Physique des Particules, IN2P3/CNRS et Universit\'e de Savoie, F-74941 Annecy-Le-Vieux, France }
\author{J.~Garra~Tico}
\author{E.~Grauges}
\affiliation{Universitat de Barcelona, Facultat de Fisica, Departament ECM, E-08028 Barcelona, Spain }
% ...
\collaboration{The \babar\ Collaboration}
\noaffiliation

\begin{abstract}
We measure branching fractions and integrated rate
asymmetries for the rare decays $\kll$,
where $\ell^+\ell^-$ is either $e^+e^-$ or $\mu^+\mu^-$,
using a sample of 384 million $\BB$ events collected with
the \babar\, detector at the \pep2\ $\epem$ collider.
We find no evidence for direct $\CP$ or lepton-flavor
asymmetries. However, for dilepton masses
below the $\jpsi$ resonance, we find evidence for
unexpectedly large isospin asymmetries in both
$\modekavgll$ and $\modekstll$\, which differ respectively
by $3.2\sigma$ and $2.7\sigma$, including systematic
uncertainties, from the Standard Model expectations.
\end{abstract}



\maketitle

\pagestyle{plain}


The decays $\kll$, where $\ellell$ is either $\epem$ or $\mumu$,
arise from flavor-changing neutral current processes
that are forbidden at tree level in the Standard Model (SM).
The lowest-order SM processes contributing to these decays
are a $W^+W^-$ box diagram, and the radiative photon and electroweak $Z$
penguin diagrams~\cite{newbabar}. Their amplitudes are expressed
in terms of hadronic form factors and effective Wilson coefficients
$C^{\mathrm eff}_{7}$, $C^{\mathrm eff}_{9}$ and $C^{\mathrm eff}_{10}$,
representing the electromagnetic penguin diagram, and
the vector part and the axial-vector part of the $Z$ penguin and
$W^+W^-$ box diagrams, respectively~\cite{Buchalla}.
New physics contributions may enter the penguin and box diagrams
at the same order as the SM diagrams, modifying the
Wilson coefficients from their SM expectations~\cite{Ali:2002jg}.

We report results herein on exclusive branching fractions, direct $\CP$ asymmetries,
the ratio of rates to di-muon and di-electron final states,
and isospin asymmetries, measured in two regions of dilepton mass squared
chosen to exclude the region of the $\jpsi$ resonance: a low $q^2$ region
$0.1 < q^2 \equiv m^2_{\ell\ell} < 7.02 \gevcccc$ and a high $q^2$ region
$q^2 > 10.24 \gevcccc$. We also present results for the two regions combined.
The $\psi(2S)$ resonance is removed from the high $q^2$ region
by vetoing events with $12.96 < q^2 < 14.06 \gevcccc$.
For $K^* \epem$ final states, we also report results
in extended low and extended combined $q^2$ regions including events $q^2 < 0.1 \gevcccc$,
where there is an enhanced coupling to the photonic penguin amplitude
unique to this mode. Recent \babar\, results on angular observables using the same dataset
and similar event selection as is used here are reported in~\cite{newbabar}.

The $\modekavgll$ branching fraction is predicted
to be $(0.35 \pm 0.12) \times 10^{-6}$,
while $\modekstll$ for $q^2 > 0.1 \gevcccc$ is expected
to be roughly three times larger at $(1.19 \pm 0.39) \times 10^{-6}$~\cite{Ali:2002jg}.
The $\sim 30\%$ uncertainties are due to lack of knowledge
about the form factors that model the hadronic effects in
the $\B \to K$ and $B \to K^{*}$ transitions. Thus, measurements
of decay rates to exclusive final states are
less suited to searches for new physics than rate asymmetries,
where many theory uncertainties cancel~\cite{Kruger:1999xa}.

The direct $\CP$ asymmetry
\begin{eqnarray}
A_{\CP}^{\Kmaybestar} \equiv
\frac
{{\cal B}(\overline{B} \rightarrow \overline{K}^{(*)}\ellell) - {\cal B}(B \rightarrow K^{(*)}\ellell)}
{{\cal B}(\overline{B} \rightarrow \overline{K}^{(*)}\ellell) + {\cal B}(B \rightarrow K^{(*)}\ellell)}
\end{eqnarray}
is expected to be $O(10^{-3})$ in the SM, but new physics at the electroweak
scale could produce a significant enhancement~\cite{Bobeth:2008hp}.

The ratio of rates to di-muon and di-electron final states
\begin{eqnarray}
R_{\Kmaybestar} \equiv
\frac
{{\cal B}(\kmaybemm)}
{{\cal B}(\kmaybeee)}
\end{eqnarray}
is unity in the SM to within a few percent~\cite{Hiller:2003js}.
In two-Higgs-doublet models, including supersymmetry,
these ratios are sensitive to the presence of a neutral
Higgs boson, which might, at large $\tan \beta$,
increase $R_{\Kmaybestar}$ by $\sim 10\%$~\cite{Yan:2000dc}.
In the region $q^2 < (2m_{\mu})^2$, where
only the $\epem$ modes are allowed, there is a large
enhancement of $\modekstee$ due to a $1/q^2$ scaling of the
photon penguin.
The expected SM value of $R_{K^*}$ including this region is
0.75~\cite{Hiller:2003js}, and we fit the $K^*$ dataset over
the extended combined and extended low $q^2$ regions in order
to test this prediction.

The \CP-averaged isospin asymmetry
\begin{eqnarray}
A^{K^{(*)}}_{I} \equiv
\frac
{{\cal B}(\Bz \to K^{(*)0}\ellell) - r {\cal B}(\B^{\pm} \to K^{(*)\pm}\ellell)}
{{\cal B}(\Bz \to K^{(*)0}\ellell) + r {\cal B}(\B^{\pm} \to K^{(*)\pm}\ellell)}
\end{eqnarray}
\noindent
where $r = \tau_0/\tau_+=1/(1.07\pm 0.01)$ is the ratio of
the $B^0$ and $B^+$ lifetimes~\cite{hfag},
has a SM expectation of $+6-13\%$ as $q^2 \rightarrow 0 \gevcccc$~\cite{Feldmann:2002iw}.
This is consistent with the measured asymmetry
of 3$\pm$3\% in $B\to K^*\gamma$~\cite{hfag}.
A calculation of the predicted $K^{*+}$ and $K^{*0}$
rates integrated over the low $q^2$
region gives $A^{K^{*}}_{I} = -0.005 \pm 0.020$~\cite{beneke05,Feldmann:ckm2008}.
In the high $q^2$ region, contributions from charmonium states
may provide an additional source of isospin asymmetry, although
the measured asymmetry in $\jpsi\Kmaybestar$ is at most a few percent~\cite{hfag}.

We use a data sample of $384$ million $\BB$ pairs
collected at the $\FourS$ resonance
with the \babar\ detector~\cite{BaBarDetector} at
the \pep2\ asymmetric-energy $\epem$ collider at SLAC.
Our selection of charged and neutral particle candidates, as
well as reconstruction of $\piz$, $\KS$ and $K^{*}$ candidates,
are described at~\cite{newbabar}.
We reconstruct signal events in ten separate final states
containing an $\epem$ or $\mumu$ pair, and a $\KS(\to \pip\pim)$, $\Kp$,
or $\Kstar(892)$ candidate with an invariant mass $0.82 < M(K\pi) < 0.97 \gevcc$.
We reconstruct $K^{*0}$ candidates in the
final state $K^+\pi^-$, and $K^{*+}$ candidates in the final states
$K^+\pi^0$ and $\KS\pi^+$ (charge conjugation is implied throughout
except as explicitly noted).
We also study final states $K^{(*)}h^{\pm}\mu^{\mp}$,
where $h$ is a track with no particle identification requirement applied,
to characterize backgrounds from hadrons misidentified as muons.

$B\to \Kmaybestar \ell^+\ell^-$ decays are reconstructed using the kinematic
variables $\mes=\sqrt{s/4 -p^{*2}_B}$ and
$\Delta E = E_B^* - \sqrt{s}/2$, where $p^*_B$ and $E_B^*$ are
the $B$ momentum and energy in the $\Upsilon(4S)$ center-of-mass (CM) frame,
and $\sqrt{s}$ is the total CM energy.
We define a fit region $\mes > 5.2 \gevcc$, with
$-0.07<\Delta E<0.04$ ($-0.04<\Delta E<0.04$) $\gev$ for
$e^+e^-$ ($\mu^+\mu^-$) final states in the low and extended low $q^2$ region, and
$-0.08<\Delta E<0.05$ ($-0.05<\Delta E<0.05$) $\gev$ for high $q^2$.

The main backgrounds arise from random combinations of
leptons from semileptonic $B$ and $D$ decays, which are
suppressed through the use of neural networks (NN) whose
construction is described in detail in~\cite{newbabar}.
For each of the ten final states we use separate NN optimized
to suppress either continuum or $B\Bbar$ backgrounds
in the low, extended low or high $q^2$ regions.
We use simulated samples of signal and background events in
the construction of the NN, and assume rates consistent with
accepted values~\cite{hfag}.

There is a further background contribution from
$B \to D(\rightarrow \Kmaybestar \pi) \pi$ decays,
where both pions are misidentified as leptons.
The pion misidentification rates are 2-3\% for muons and
$<$0.1\% for electrons, so this background is only significant
in the $\mu^+\mu^-$ final states.
We veto these events by assigning the pion mass to a muon candidate,
and requiring the invariant mass of the hypothetical $\Kmaybestar\pi$ system
to be outside the range 1.84-1.90$\gevcc$.
After all the above selections have been applied, the final reconstruction
efficiency for signal events varies from 3.5\% for $K^+\pi^0\mu^+\mu^-$
for the combined $q^2$ region, to 22\% for $K^+\pi^-e^+e^-$ in the high $q^2$ region.

We perform unbinned maximum likelihood fits to $\mes$ distributions
to obtain signal and background yields.
We use an ARGUS shape~\cite{ArgusShape}
to describe the combinatorial background, allowing
the shape parameter to float in the fits.
For the signal, we use a fixed Gaussian shape unique to each final state,
with mean and width determined from fits to the analogous final states in
the vetoed $J/\psi\Kmaybestar$ events.
We account for a small residual contribution
from misidentified hadrons by constructing a
probability density function (pdf) using
$\Kmaybestar h^{\pm}\mu^{\mp}$ events weighted by the probability for the
$h^{\pm}$ to be misidentified as a muon.
We also account for background events that peak in the $\mes$ signal region,
arising from charmonium events that escape the veto, and
for contributions from misreconstructed signal events.
We test our fits in each final state using the large
samples of vetoed $J/\psi \Kmaybestar$ and $\psitwos \Kmaybestar$ events, and
find that all the branching fractions are in good agreement with
accepted values~\cite{PDG}. We perform simultaneous fits for
$A_{\CP}^{\Kmaybestar}$, $R_{\Kmaybestar}$ and $A^{K^{(*)}}_{I}$ summed
over all the signal modes that contribute to the particular measurement.

We estimate the statistical significance of our fits
by generating ensembles of 1000 datasets for each of the ten final states
in each $q^2$ region of interest, and fitting each dataset with the
full fit model described above.
These tests also confirm the unbiased nature and
proper error scaling of our fit methodology.

For the total $\modekavgll$ and $\modekstll$ branching fractions
averaged assuming isospin and lepton-flavor symmetry, we measure
$(0.394_{-0.069}^{+0.073}$ $\pm0.020) \times 10^{-6}$ and
$(1.11_{-0.18}^{+0.19}$    $\pm0.07)  \times 10^{-6}$, respectively,
where the first uncertainty is statistical and the second is systematic.
Complete branching fraction results in all final states and $q^2$ regions,
along with the statistical significance of each measurement
and frequentist upper limits for measurements with $<4\sigma$ statistical significance,
are available online~\cite{epaps}. All results are in good agreement with
previous measurements~\cite{hfag}.

Table \ref{tab:acpressys} summarizes the results for $A_{\CP}^{\Kmaybestar}$.
In the fits to the separate $B$ and $\Bbar$ datasets in charge-conjugate
final states, we assume a common background ARGUS shape parameter.
Our final results are consistent with the SM
expectation of negligible direct $\CP$ asymmetry.
Table~\ref{tab:emuressys} shows the results for $R_K$ and $R_{K^*}$,
which are also consistent with the SM expectations.

\begin{table}[b!]
\centering
\caption{$A_{\CP}^{\Kmaybestar}$ results in each relevant $q^2$ region. The
uncertainties are statistical and systematic, respectively.}
\label{tab:acpressys}
{\footnotesize
\begin{tabular}{lccc}
\hline \hline
Mode & combined $q^2$ & low $q^2$ & high $q^2$
\\ \hline
$\modekllshort$ & $-0.18_{-0.18}^{+0.18}\pm0.01$ & $-0.18_{-0.19}^{+0.19}\pm0.01$ & $-0.09_{-0.39}^{+0.36}\pm0.02$   \\
$\modekstkllshort$ & $0.02_{-0.20}^{+0.20}\pm0.02$ & $-0.23_{-0.38}^{+0.38}\pm0.02$ & $0.17_{-0.24}^{+0.24}\pm0.02$  \\
$\modekstksllshort$ & $0.01_{-0.24}^{+0.26}\pm0.02$ & $0.10_{-0.24}^{+0.25}\pm0.02$ & $-0.18_{-0.55}^{+0.45}\pm0.04$ \\
$\modekstllshort$ & $0.01_{-0.15}^{+0.16}\pm0.01$ & $0.01_{-0.20}^{+0.21}\pm0.01$ & $0.09_{-0.21}^{+0.21}\pm0.02$    \\
\hline \hline
\end{tabular}
}
\end{table}

\begin{table}
\caption{$R_{\Kmaybestar}$ results in each $q^2$ region.
The extended (``ext.'') regions are relevant only for $R_{K^*}$.
The uncertainties are statistical and systematic, respectively.}
\centering
\begin{tabular}{lcc}
\hline \hline
$q^2$ Region  & $R_{K^*}$                      & $R_K$
\\ \hline
combined          & $1.37_{-0.40}^{+0.53}\pm 0.09$ & $0.96_{-0.34}^{+0.44}\pm 0.05$ \\
ext. combined     & $1.10_{-0.32}^{+0.42}\pm 0.07$ & ---                            \\
low               & $1.01_{-0.44}^{+0.58}\pm 0.08$ & $0.40_{-0.23}^{+0.30}\pm 0.02$ \\
ext. low          & $0.56_{-0.23}^{+0.29}\pm 0.04$ & ---                            \\
high              & $2.15_{-0.78}^{+1.42}\pm 0.15$ & $1.06_{-0.51}^{+0.81}\pm 0.06$ \\
\hline \hline
\end{tabular}
\label{tab:emuressys}
\end{table}

\begin{figure}
\begin{center}
\includegraphics[width=0.5\textwidth]{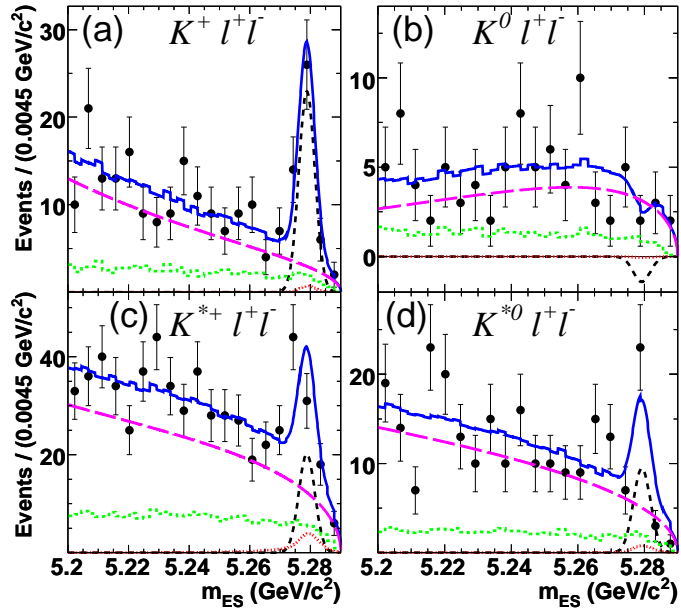}
\caption{Charged and neutral
fit projections in the low $q^2$ region.
Total fit [solid],
combinatoric background [long dash],
signal [medium dash],
hadronic background [short dash],
peaking background [dots].}
\label{fig:alliso}
\end{center}
\end{figure}

Table~\ref{tab:isoressys} shows the results for the isospin asymmetry $A_{I}^{\Kmaybestar}$.
We directly fit the data for $A_{I}^{\Kmaybestar}$ taking into account the
differing lifetimes of $\Bz$ and $\Bp$.
Figure~\ref{fig:alliso} shows the charged and neutral
low $q^2$ datasets with overlaid fit projections.
We find no significant isospin asymmetries in the high and combined $q^2$ regions,
or for $\modeksteeshort$ fits in the extended regions.
However, we find evidence for large negative asymmetries in the low $q^2$ region.

\begin{table}
\centering
\caption{$A_{I}^{\Kmaybestar}$ results in each $q^2$ region.
The uncertainties are statistical and systematic, respectively.
The last table row shows $\modeksteeshort$ results for the extended regions.}
{\footnotesize
\begin{tabular}{lcccc}
\hline \hline
Mode          & combined $q^2$  & low $q^2$ & high $q^2$
\\ \hline
$\modekavgmmshort$ & $0.13_{-0.37}^{+0.29}\pm0.04$   &  $-0.91_{\mathrm{-\infty}}^{+1.2}\pm0.18$  & $0.39_{-0.46}^{+0.35}\pm0.04$ \\
$\modekavgeeshort$ & $-0.73_{-0.50}^{+0.39}\pm0.04$  & $-1.41_{-0.69}^{+0.49} \pm 0.04$        & $0.21_{-0.41}^{+0.32}\pm0.03$ \\
$\modekavgllshort$ & $-0.37_{-0.34}^{+0.27}\pm0.04$  & $-1.43_{-0.85}^{+0.56} \pm 0.05$        & $0.28_{-0.30}^{+0.24}\pm0.03$ \\
$\modekstmmshort$  & $-0.00_{-0.26}^{+0.36}\pm0.05$  & $-0.26_{-0.34}^{+0.50} \pm0.05$         &$-0.08_{-0.27}^{+0.37}\pm0.05$ \\
$\modeksteeshort$  & $-0.20_{-0.20}^{+0.22}\pm0.03$  & $-0.66_{-0.17}^{+0.19} \pm0.02$         & $0.32_{-0.45}^{+0.75}\pm0.03$ \\
$\modekstllshort$  & $-0.12_{-0.16}^{+0.18}\pm0.04$  & $-0.56_{-0.15}^{+0.17} \pm0.03$         & $0.18_{-0.28}^{+0.36}\pm0.04$ \\ \hline
$\modeksteeshort$  & $-0.27_{-0.18}^{+0.21}\pm 0.03$ & $-0.25_{-0.18}^{+0.20} \pm 0.03$        & ---                           \\
\end{tabular}
}
\label{tab:isoressys}
\end{table}

\begin{figure}
\begin{center}
\includegraphics[width=0.4\textwidth]{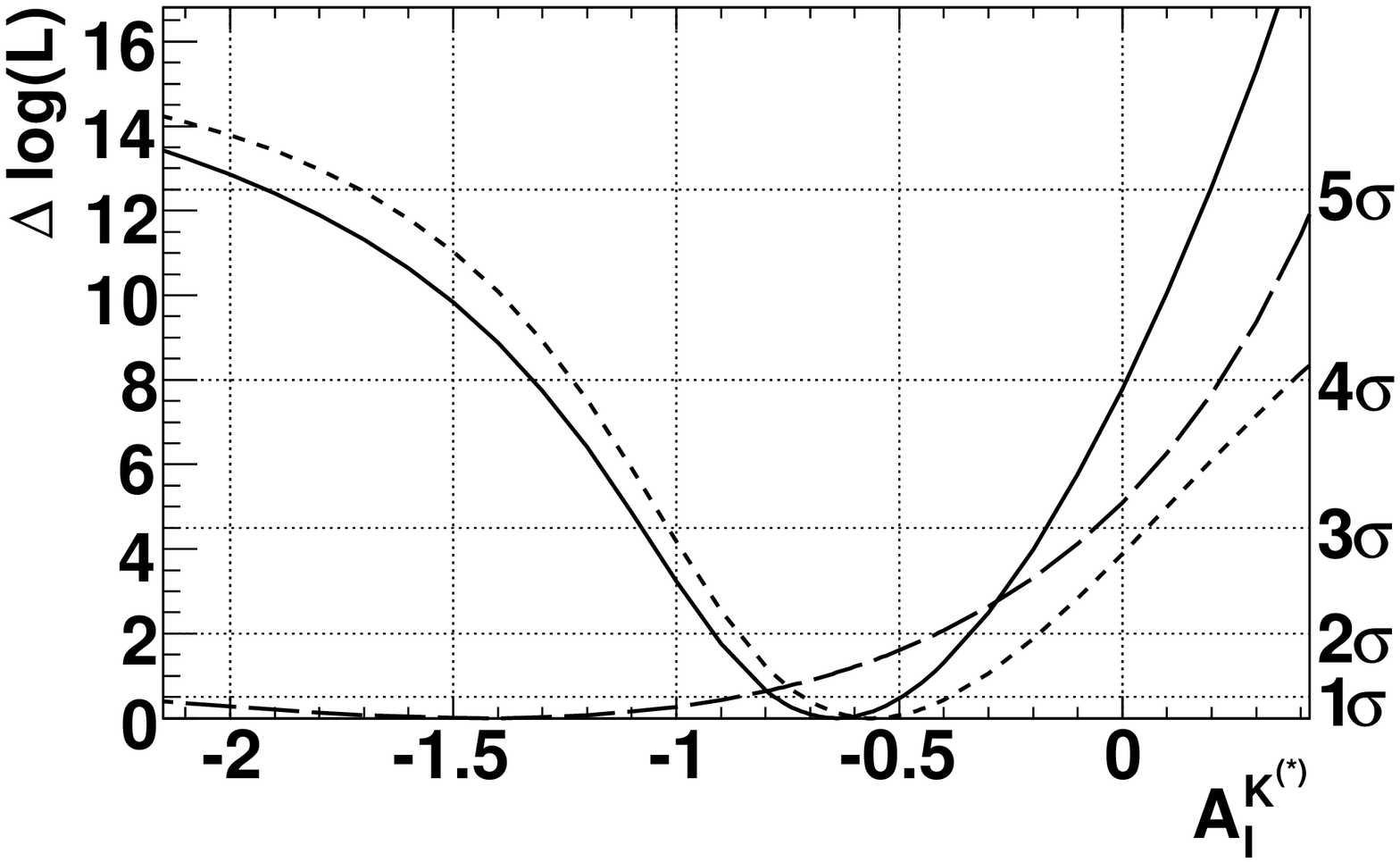}
\caption{Low $q^2$ region $A_{I}^{\Kmaybestar}$ fit likelihood curves.
$\modekavgllshort$ [long dash],
$\modekstllshort$ [short dash],
$\kllshort$ [solid].}
\label{fig:nllscans}
\end{center}
\end{figure}

We calculate the statistical significance with which a null
isospin asymmetry hypothesis is rejected using the change in
log likelihood $\sqrt{2 \Delta \ln{\cal L}}$ between the nominal
fit to the data and a fit with $A_{I}^{\Kmaybestar}=0$ fixed.
Figure~\ref{fig:nllscans} shows the likelihood curves obtained
from the $\modekavgllshort$ and $\modekstllshort$
fits. The parabolic nature of the curves in the $A_{I}^{\Kmaybestar}>-1$ region
demonstrates the essentially Gaussian nature of our fit results
in the physical region, and the right-side axis of
Figure~\ref{fig:nllscans} shows purely statistical significances
based on Gaussian coverage.
Incorporating the relatively small systematic uncertainties as
a scaling factor on the change in log likelihood, the significance
in the low $q^2$ region that $A_{I}^{\Kmaybestar}$
is different from zero is $3.2\sigma$ for $\modekavgllshort$
and $2.7\sigma$ for $\modekstllshort$.
We have verified these confidence intervals by performing fits to ensembles
of simulated datasets generated with $A_{I}^{\Kmaybestar}=0$ fixed, and we find
frequentist coverage consistent with the $\Delta \ln{\cal L}$ calculations.
The highly negative $A_{I}^{\Kmaybestar}$ values for both $\modekavgllshort$ and $\modekstllshort$
at low $q^2$ suggest that this asymmetry may be insensitive to the hadronic final state,
and so we sum the likelihood curves as shown in Figure~\ref{fig:nllscans} and obtain
$A_{I}^{\Kmaybestar} = -0.64^{+0.15}_{-0.14} \pm 0.03$. Including systematics, this is a $3.9 \sigma$
difference from a null $A_{I}^{\Kmaybestar}$ hypothesis.

We consider systematic uncertainties associated with
reconstruction efficiencies;
hadronic background parameterization in di-muon final states;
peaking background contributions obtained from simulated events;
and possible $\CP$, lepton flavor and isospin asymmetries in the
background pdfs. We quantify the efficiency systematics using the vetoed
$\jpsi\Kmaybestar$ samples. These include charged track, $\piz$, and \KS
reconstruction, particle identification, NN selection,
and the $\DeltaE$ and $\K^{*}$ mass selections.
The largest contributions to the systematic uncertainties on the rates are
particle identification, the characterization of the
hadronic background and the signal $\mes$ pdf shape.
All of these cancel at least partially in the rate asymmetries,
and the final systematic uncertainties are small compared to the
statistical ones.

We perform several additional checks of
effects that might cause a bias in our final results.
We vary the parameterization of the hadronic background pdfs,
and of the random combinatorial background ARGUS shapes in the low
$q^2$ region, to test the robustness of the large $A_{I}^{\Kmaybestar}$ asymmetries.
We remove all the NN selections, and perform
separate fits to the two $\Kstarp$ final states, and
observe no significant variation in the $A_{I}^{\Kmaybestar}$ results.
To understand if an isospin asymmetry might be induced by
the combinatorial background, we compare data
and simulated background events within a larger region
$\left|\DeltaE\right|<0.25 \gev$ outside our $\DeltaE$ selection
window and in the $5.2 < \mes < 5.27 \gevcc$ region.
We find that the numbers of simulated and data events
in this larger region agree well. No signal isospin asymmetry
is found using simulated events within the fit region.

In summary, we have measured branching fractions, and studied direct $\CP$ violation,
ratios of rates to di-muon and di-electron final
states, and isospin asymmetries
in the rare decays $\kll$. Our branching fraction results agree
with both SM predictions and previous measurements.
Our results for the direct $\CP$ asymmetries and lepton-flavor rate
ratios are in good agreement with their respective SM predictions of zero
and one. The isospin asymmetries in the high and combined $q^2$ regions are
consistent with zero, but in the low $q^2$ region
in both $\modekavgll$ and $\modekstll$ we measure large negative asymmetries
that are each about $3\sigma$ different from zero,
including systematic uncertainties. Combining these results,
we obtain $A_{I}^{\Kmaybestar} = -0.64^{+0.15}_{-0.14} \pm 0.03$,
with a $3.9 \sigma$ difference (including systematics) from $A_{I}^{\Kmaybestar}=0$.
Such large negative asymmetries are unexpected in the SM,
which predicts essentially no isospin asymmetry integrated
over our low $q^2$ region and, as $q^2 \rightarrow 0$, an
asymmetry of $\sim +10\%$, opposite in sign to our observation
in the low $q^2$ region.

We are grateful for the excellent luminosity and machine conditions
provided by our \pep2\ colleagues,
and for the substantial dedicated effort from
the computing organizations that support \babar.
The collaborating institutions wish to thank
SLAC for its support and kind hospitality.
This work is supported by
DOE
and NSF (USA),
NSERC (Canada),
CEA and
CNRS-IN2P3
(France),
BMBF and DFG
(Germany),
INFN (Italy),
FOM (The Netherlands),
NFR (Norway),
MES (Russia),
MEC (Spain), and
STFC (United Kingdom).
Individuals have received support from the
Marie Curie EIF (European Union) and
the A.~P.~Sloan Foundation.

\clearpage


\begin{thebibliography}{99}
\def\mpl #1 #2 #3 {Mod.~Phys.~Lett.~{\bf#1},\ #2 (#3)}
\def\npb  #1 #2 #3 {Nucl.~Phys.~B~{\bf#1},\ #2 (#3)}
\def\plb  #1 #2 #3 {Phys.~Lett.~B~{\bf#1},\ #2 (#3)}
\def\pr   #1 #2 #3 {Phys.~Rep.~{\bf#1},\ #2 (#3)}
\def\prd  #1 #2 #3 {Phys.~Rev.~D~{\bf#1},\ #2 (#3)}
\def\prl  #1 #2 #3 {Phys.~Rev.~Lett.~{\bf#1},\ #2 (#3)}
\def\RMP  #1 #2 #3 {Rev.~Mod.~Phys.~{\bf#1},\ #2 (#3)}
\def\zpc  #1 #2 #3 {Z.~Phys.~C~{\bf#1},\ #2 (#3)}
\def\nim  #1 #2 #3 {Nucl.~Instrum.~Methods~{\bf#1},\ #2 (#3)}
\def\nima  #1 #2 #3 {Nucl.~Instrum.~Methods~A~{\bf#1},\ #2 (#3)}
\def\epjc #1 #2 #3 {Eur.~Phys.~J.~C~{\bf#1},\ #2 (#3)}
\def\rmp #1 #2 #3 {Rev.~Mod.~Phys.~{\bf#1},\ #2 (#3)}
\def\npbps #1 #2 #3 {Nucl.~Phys.~B.~proc.~suppl~{\bf#1},\ #2 (#3)}
\def\progtp #1 #2 #3 {Prog.~Theo.~Phys~{\bf#1},\ #2 (#3)}
\def\etal{{\it et al.}}


\bibitem{newbabar}
  B.~Aubert {\it et al.}  [\babar\ Collaboration],
  arXiv:0804.4412 (2008), submitted to PRD-RC.
\bibitem{Buchalla}
  G.~Buchalla, A.~J.~Buras and M.~E.~Lautenbacher,
  Rev.\ Mod.\ Phys.\  {\bf 68}, 1125 (1996).
\bibitem{Ali:2002jg}
  A.~Ali, E.~Lunghi, C.~Greub and G.~Hiller,
  Phys.\ Rev.\  D {\bf 66}, 034002 (2002).
\bibitem{Kruger:1999xa}
  F.~Kruger, L.~M.~Sehgal, N.~Sinha and R.~Sinha,
  Phys.\ Rev.\  D {\bf 61}, 114028 (2000)
  [Erratum-ibid.\  D {\bf 63}, 019901 (2001)].
\bibitem{Bobeth:2008hp}
  C.~Bobeth, G.~Hiller and G.~Piranishvili,
  arXiv:0805.2525 (2008).
\bibitem{Hiller:2003js}
  G.~Hiller and F.~Kruger,
  Phys.\ Rev.\  D {\bf 69}, 074020 (2004).
\bibitem{Yan:2000dc}
  Q.~S.~Yan, C.~S.~Huang, W.~Liao and S.~H.~Zhu,
  Phys.\ Rev.\  D {\bf 62}, 094023 (2000).
\bibitem{hfag}
  Heavy Flavor Averaging Group,
  E.~Barberio {\it et al.}, arXiv:0704.3575 (2007).
\bibitem{Feldmann:2002iw}
  T.~Feldmann and J.~Matias,
  JHEP {\bf 0301}, 074 (2003).
\bibitem{beneke05}
 M.~Beneke, T.~Feldmann and D.~Seidel
 Eur. Phys. J. \textbf{C41}, 173 (2005).
\bibitem{Feldmann:ckm2008}
  T.~Feldmann, 5th Workshop on the CKM Unitary Triangle, Rome (2008).
\bibitem{BaBarDetector}
  B.~Aubert {\it et al.}  [\babar\ Collaboration],
  Nucl.\ Instrum.\ Meth.\  A {\bf 479}, 1 (2002).
\bibitem{ArgusShape}
H.~Albrecht {\it et al.}  [ARGUS Collaboration],
  Z.\ Phys.\  C {\bf 48}, 543 (1990).
\bibitem{PDG}
W.~M.~Yao {\it et al.}  [Particle Data Group],
  J.\ Phys.\ G {\bf 33} (2006) 1.
\bibitem{epaps}
See EPAPS Document No. E-PRLTAO-102-060910 for
branching fraction results to accompany this
article. For more information on EPAPS,
see http://www.aip.org/pubservs/epaps.html.

\end{thebibliography}
\end{document}



\title{
        {\mathversion{bold}
         $\kll$ Branching Fraction Results}
}

\pagestyle{plain}


Table~\ref{tab:bfsummary} shows the total and partial branching
fractions in $\kll$ modes, along with statistical and systematic
errors, and the statistical significance of each measurement.
To calculate significance, the data are refit with a null signal
hypothesis and the change in $\ln{\cal L}$ with the nominal fit
is used to determine the significance.
For measurements with a statistical significance less than $4.0\sigma$,
we fit ensembles of toy datasets derived from our observations in the data
to compute frequentist upper limits at the 90\% confidence level, in which
only fit results which give a physical signal yield are used.
The combined $q^2$ results are scaled to account
for the regions of the total dilepton mass distribution which are
vetoed and not included in the fits.

\begin{table}[h!]
\centering
\caption{Branching fractions by mode and $q^2$ region with statistical and systematic errors, respectively.
The statistical significance of each measurement is given parenthetically. 90\% confidence level upper limits are
provided for measurements with $<4\sigma$ statistical significance.}
\label{tab:bfsummary}
\begin{tabular}{lcccccc}
\\ \hline \hline \\
Mode                & combined $q^2$ $(10^{-6})$                     & 90\% UL & low $q^2$ $(10^{-6})$                               & 90\% UL & high $q^2$ $(10^{-6})$                             & 90\% UL \\ \hline \\
$\modekzmmshort$    & $0.49_{-0.25}^{+0.29}$ $\pm0.03$ ($2.2\sigma$) & 0.99    & $0.01_{-0.13}^{+0.15}$     $\pm0.01 $ ($0.0\sigma$) & 0.31    & $0.25_{ -0.14 }^{+0.16 }$ $\pm0.02 $ ($1.9\sigma$) & 0.53    \\
$\modetwoshort$     & $0.41_{-0.15}^{+0.16}$ $\pm0.02$ ($3.0\sigma$) & 0.62    & $0.123_{-0.071}^{+0.082}$  $\pm0.007$ ($1.8\sigma$) & 0.25    & $0.116_{-0.072}^{+0.079}$ $\pm0.010$ ($1.7\sigma$) & 0.24    \\
$\modekzeeshort$    & $0.08_{-0.12}^{+0.15}$ $\pm0.01$ ($0.6\sigma$) & 0.36    & $-0.049_{-0.049}^{+0.061}$ $\pm0.004$ ($0.0\sigma$) & 0.10    & $0.177_{-0.091}^{+0.118}$ $\pm0.011$ ($2.3\sigma$) & 0.38    \\
$\modefourshort$    & $0.51_{-0.11}^{+0.12}$ $\pm0.02$ ($5.8\sigma$) & --      & $0.308_{-0.062}^{+0.070}$  $\pm0.013$ ($6.7\sigma$) & --      & $0.125_{-0.055}^{+0.062}$ $\pm0.006$ ($2.5\sigma$) & 0.21    \\
\\ \hline \\
$\modekstksmmshort$ & $1.46_{-0.75}^{+0.79}$ $\pm0.12$ ($2.0\sigma$) & 2.55    & $0.75_{-0.45}^{+0.49}$     $\pm0.07$  ($1.7\sigma$) & 1.48    & $0.78_{-0.40}^{+0.43}$    $\pm0.06$  ($2.1\sigma$) & 1.36    \\
$\modekstzmmshort$  & $1.35_{-0.37}^{+0.40}$ $\pm0.10$ ($4.1\sigma$) & --      & $0.41_{-0.21}^{+0.24}$     $\pm0.04$  ($2.2\sigma$) & 0.83    & $0.62_{-0.19}^{+0.21}$    $\pm0.05$  ($3.8\sigma$) & 0.97    \\
$\modekstkseeshort$ & $1.38_{-0.42}^{+0.47}$ $\pm0.08$ ($3.7\sigma$) & 1.97    & $1.06_{-0.28}^{+0.31}$     $\pm0.07$  ($4.7\sigma$) & --      & $0.19_{-0.21}^{+0.23}$    $\pm0.01$  ($0.9\sigma$) & 0.53    \\
$\modekstzeeshort$  & $0.86_{-0.24}^{+0.26}$ $\pm0.05$ ($4.2\sigma$) & --      & $0.20_{-0.11}^{+0.12}$     $\pm0.01$  ($2.1\sigma$) & 0.42    & $0.35_{-0.13}^{+0.15}$    $\pm0.02$  ($3.0\sigma$) & 0.60    \\
\\\hline \hline
\end{tabular}
\end{table}

In addition to the above results, we have also performed fits in
extended $q^2$ regions for $K^{*}$ di-electron final states, where
a significant rate enhancement compared to that above $q^2 > 0.1 \gevcccc$ is expected.
Table~\ref{tab:bfsummarypolesys} shows the total and partial branching fractions
for these modes for extended $q^2$ regions.

\begin{table}[h!]
\centering
\caption{Branching fractions for the extended $q^2$ regions with statistical and systematic errors, respectively.}
\label{tab:bfsummarypolesys}
\begin{tabular}{lcc}
\\ \hline \hline \\
Mode            & combined $q^2$ $(10^{-6})$    & low $q^2$ $(10^{-6})$         \\ \hline \\
$\modekstksee$  & $1.90_{-0.55}^{+0.59}\pm0.11$ & $1.32_{-0.36}^{+0.41}\pm0.09$ \\
$\modekstzee$   & $1.02_{-0.28}^{+0.30}\pm0.06$ & $0.73_{-0.19}^{+0.22}\pm0.04$ \\
\\\hline \hline
\end{tabular}
\end{table}

We also perform fits combining various hadronic and di-lepton
final states which are then averaged assuming
isospin and/or lepton-flavor symmetry.
Table~\ref{tab:bfsummarycombined} shows branching fractions for the combined modes
including statistical and systematic errors, statistical significance and upper limits
for measurements with $<4\sigma$ statistical significance.

\begin{table}[h!]
\centering
\caption{
Combined mode branching fractions by $q^2$ region with statistical and systematic errors, respectively.
The statistical significance of each measurement is given parenthetically.
90\% confidence level upper limits are provided for measurements with $<4\sigma$ statistical significance.}
\label{tab:bfsummarycombined}
\begin{tabular}{lcccccc}
\\ \hline \hline \\
Mode                & combined $q^2$ $(10^{-6})$                         & 90\% UL & low $q^2$ $(10^{-6})$                                & 90\% UL & high $q^2$ $(10^{-6})$                             & 90\% UL \\ \hline \\
$\modekzllshort$    & $0.21_{-0.13}^{+0.15}$    $\pm0.02 $ $(1.7\sigma)$ & 0.45    & $-0.041_{-0.047}^{+0.058}$ $\pm0.004$  $(0.0\sigma)$ & 0.09    & $0.202_{-0.081}^{+0.096}$ $\pm0.013$ $(3.0\sigma)$ & 0.36    \\
$\modekllshort$     & $0.476_{-0.086}^{+0.092}$ $\pm0.022$ $(6.5\sigma)$ & --      & $ 0.250_{-0.047}^{+0.052}$ $\pm0.010$  $(6.7\sigma)$ & --      & $0.122_{-0.044}^{+0.048}$ $\pm0.006$ $(3.0\sigma)$ & 0.19    \\
$\modekavgmmshort$  & $0.41_{-0.12}^{+0.13}$    $\pm0.02 $ $(4.7\sigma)$ & --      & $0.096_{-0.059}^{+0.066}$  $\pm0.006$  $(2.1\sigma)$ & 0.21    & $0.139_{-0.062}^{+0.067}$ $\pm0.008$ $(2.9\sigma)$ & 0.25    \\
$\modekavgeeshort$  & $0.388_{-0.083}^{+0.090}$ $\pm0.020$ $(5.6\sigma)$ & --      & $0.217_{-0.046}^{+0.051}$  $\pm0.010$  $(5.9\sigma)$ & --      & $0.132_{-0.045}^{+0.050}$ $\pm0.007$ $(3.5\sigma)$ & 0.21    \\
$\modekavgllshort$  & $0.394_{-0.069}^{+0.073}$ $\pm0.020$ $(7.3\sigma)$ & --      & $0.181_{-0.036}^{+0.039}$  $\pm0.008$  $(6.1\sigma)$ & --      & $0.135_{-0.037}^{+0.040}$ $\pm0.007$ $(4.5\sigma)$ & --      \\
\\ \hline \\
$\modekstkllshort$  & $1.03_{-0.21}^{+0.22}$    $\pm0.07 $ $(5.8\sigma)$ & --      & $ 0.257_{-0.098}^{+0.110}$ $\pm0.02 $  $(2.9\sigma)$ & 0.44    & $ 0.46_{-0.11 }^{+0.12 }$ $\pm0.03 $ $(4.7\sigma)$ & --      \\
$\modekstksllshort$ & $1.40_{-0.37}^{+0.40}$    $\pm0.09 $ $(4.2\sigma)$ & --      & $  0.98_{-0.24 }^{+0.26 }$ $\pm0.06 $  $(4.9\sigma)$ & --      & $ 0.34_{-0.19 }^{+0.21 }$ $\pm0.02 $ $(1.8\sigma)$ & 0.63    \\
$\modekstmmshort$   & $1.35_{-0.33}^{+0.35}$    $\pm0.10 $ $(5.3\sigma)$ & --      & $0.47_{-0.19}^{+0.21}$     $\pm0.04 $  $(3.2\sigma)$ & 0.84    & $0.65_{-0.17}^{+0.18}$    $\pm0.05 $ $(4.7\sigma)$ & --      \\
$\modeksteeshort$   & $0.99_{-0.21}^{+0.23}$    $\pm0.06 $ $(5.6\sigma)$ & --      & $0.42_{-0.12}^{+0.13}$     $\pm0.03 $  $(4.3\sigma)$ & --      & $ 0.30_{-0.11 }^{+0.12 }$ $\pm0.02 $ $(3.1\sigma)$ & 0.49    \\
$\modekstllshort$   & $1.11_{-0.18}^{+0.19}$    $\pm0.07 $ $(7.7\sigma)$ & --      & $0.43_{-0.10}^{+0.11}$     $\pm0.03 $  $(5.3\sigma)$ & --      & $0.42_{-0.10}^{+0.10}$    $\pm0.03 $ $(5.3\sigma)$ & --      \\
\\\hline \hline
\end{tabular}
\end{table}